\documentclass[11pt]{article}

\usepackage[margin=1in]{geometry}
\usepackage{setspace}
\usepackage[T1]{fontenc}
\usepackage[utf8]{inputenc}
\usepackage{lmodern}
\usepackage{textcomp}
\usepackage{microtype}

\usepackage{amsmath,amssymb,amsthm}
\usepackage{bm}
\usepackage{array}
\usepackage{booktabs}
\usepackage{calc}
\usepackage{caption}
\usepackage{subcaption}
\usepackage{enumitem}
\usepackage{float}
\usepackage{graphicx}
\usepackage{longtable}
\usepackage{multirow}
\usepackage{multicol}
\usepackage{threeparttable}
\usepackage[boxed]{algorithm2e}
\usepackage{xcolor}
\usepackage{comment}

\makeatletter
\def\maxwidth{\ifdim\Gin@nat@width>\linewidth\linewidth\else\Gin@nat@width\fi}
\def\maxheight{\ifdim\Gin@nat@height>\textheight\textheight\else\Gin@nat@height\fi}
\makeatother
\setkeys{Gin}{width=\maxwidth,height=\maxheight,keepaspectratio}

\makeatletter
\@ifundefined{c@theorem}{\newtheorem{theorem}{Theorem}}{}
\@ifundefined{c@lemma}{\newtheorem{lemma}{Lemma}}{}
\@ifundefined{c@assumption}{\newtheorem{assumption}{Assumption}}{}
\@ifundefined{c@remark}{\theoremstyle{remark}\newtheorem{remark}{Remark}}{}
\makeatother

\usepackage{marquis_general}
\usepackage{MRL_notation}
\usepackage{xr}
\newcommand{\csection}[1]
    {\begin{center}
        \stepcounter{section}
        {\bf\large\arabic{section}. #1}
    \end{center}
}

\newcommand{\csubsection}[1]{
\begin{center}
\stepcounter{subsection}
{\it\arabic{section}.\arabic{subsection}. #1}
\end{center}
}

\usepackage[round,authoryear]{natbib}
\usepackage{xurl}
\usepackage[colorlinks=true,linkcolor=blue,citecolor=blue,urlcolor=blue]{hyperref}
\usepackage{bookmark}

\hypersetup{
  pdftitle={Martingale R-learner:
Estimating Time-varying Heterogeneous Treatment Effects for Time-to-event Outcomes},
  pdfauthor={Jue Hou; Yuchen Qi; Ronghui Xu},
  pdfkeywords={aging study; functional score; machine learning; Neyman orthogonality; structural nested failure time model.},
  pdfcreator={LaTeX}
}

\title{Martingale R-learner:
Estimating Time-varying Heterogeneous Treatment Effects for Time-to-event Outcomes}

\author{
Jue Hou\\
\small Division of Biostatistics and Health Data Science, School of Public Health, University of Minnesota
\small \texttt{hou00123@umn.edu}
Yuchen Qi\\
\small Biostatistics and Bioinformatics, Herbert Wertheim School of Public Health and Human Longevity Science, University of California, San Diego\\
\and
Ronghui Xu\\
\small Biostatistics and Bioinformatics, Herbert Wertheim School of Public Health and Human Longevity Science,
\& Department of Mathematics, \& Halicioglu Data Science Institute
University of California, San Diego
\and
}

\date{}

\begin{document}
% \maketitle
\begin{center}
  {\bf\Large Martingale R-learner:
Estimating Time-varying Heterogeneous Treatment Effects for Time-to-event Outcomes}

  Jue Hou$^{1}$, Yuchen Qi$^{2}$, Ronghui Xu$^{2,3,4}$

{\it\small
$^1$ Division of Biostatistics and Health Data Science, School of Public Health, University of Minnesota;\\
$^2$ Biostatistics and Bioinformatics, Herbert Wertheim School of Public Health and Human Longevity Science,
University of California, San Diego; \\
$^3$  Department of Mathematics,
University of California, San Diego; \\
$^4$ Halicioglu Data Science Institute
University of California, San Diego; \\
}

\end{center}

\begin{singlespace}
\begin{abstract}
Biological research and clinical evidence suggest that treatment response may vary substantially along characteristics, such as comorbidities, genetic variants, environmental, or socio-economic factors. Future precision medicine requires accurate assessment of heterogeneous treatment effects (HTE) to guide optimal clinical decisions at the individual level. We introduce a functional score framework that extends the traditional estimating equations for survival data to nonparametric HTE and generalize the Neyman orthogonality accordingly, thus filling a methodological as well as theoretical gap. Under the Neyman orthogonal functional score framework, we developed the martingale R-learner based on a decomposition of the conditional martingale residuals into residuals of the risk-set propensity score and the marginal martingale, thereby reducing the impact of estimation bias in HTE from nuisance models including (1) marginal survival, and (2) risk-set propensity scores. This enables leveraging advances in machine learning and incorporates flexible estimators for the nuisance functions and attaining the standard optimal nonparametric estimation rate with the oracle property. Numerical experiments demonstrated empirical performance consistent with the theory. We applied the martingale R-learner to estimate the effect of alcohol on dementia using the Honolulu-Asia Aging Study data.
\end{abstract}

\noindent\textbf{Keywords:} aging study; functional score; machine learning;
 Neyman orthogonality;
structural nested failure time model.

\end{singlespace}

\csection{Introduction}

A critical question in biomedical or epidemiological research in precision medicine is understanding each individual's response to treatment, i.e. the heterogeneous treatment effect (HTE), varying along the personal characteristics \citep{kravitz2004evidence,hamburg2010path}.
For research on  cognitive impact of alcohol consumption,
early studies have noticed  heterogeneity among age groups \citep{dufour1992alcohol}, which was confirmed in subsequent studies \citep{brumback2007effects,zanjani2013alcohol}.
Other studies have investigated the roles of characteristics, such as sex and \apoe[] genotype, as the effect modifiers for alcohol consumption on cognitive impairment \citep{sabia2010effect,abulseoud2024sex}.
As the development of cognitive impairment is often a gradual process,
cohort studies with long-term follow up, such as the Honolulu-Asia Aging Study (HAAS), are extremely valuable  for uncovering the personalized patterns of alcohol-cognition relationship \citep{p2012honolulu}.
The HAAS extended the follow-up for surviving participants of the Honolulu Heart Program, which collected  drinking data from Exams 1--3 (1965 -- 1974)  along with histories of cardiovascular diseases and risk factors, and collected additionally the longitudinal cognitive assessments from Exams 4 -- 12 (1991 -- 2012).
Previous studies using different subsets of HAAS reported inconsistent results about the impact of mid-life alcohol on later life cognitive impairment  \citep{galanis2000longitudinal, kardaun2000genotypes, chosy2022midlife},
and it has been noted that the lack of a coherent framework on HTE jointly over all patient characteristics may lead to spurious findings \citep{assmann2000subgroup}.
To fill the scientific gap, we aim to develop an appropriate statistical method for systematically estimating the HTE of mid-life heavy drinking on
cognitive impairment-free survival in the HAAS cohort.

 Considerable efforts have been put into developing advanced estimation methods for the HTE without censoring.  \cite{kunzel2019metalearners} coined the concept of ``metalearners'' and classified previous methods as T-learners and S-learners  based on if two regression models (T-learner) or a single regression model (S-learner) was involved for estimating the HTE \citep{hill2011bayesian,green2012modeling,athey2016recursive}.
 They also proposed the X-learners that construct the pseudo-outcomes of the unobservable individual treatment effect by
 imputing the outcome under the unobserved counterfactual treatment \citep{kunzel2019metalearners}.
 Under the broad metalearner framework, various machine learning methods have been applied for the underlying regression problems, including lasso \citep{imai2013estimating}, random forests \citep{wager2018estimation}, boosting \citep{powers2018some}, neural networks \citep{shalit2017estimating} and Bayesian additive regression trees \citep{hill2011bayesian, hahn2020bayesian}.

 The R-learner extends the Robinson's estimator \citep{robinson1988root,chernozhukov2018double} for the average treatment effect to HTE \citep{nie2021quasioracle}.
 By regressing the residuals of a marginal outcome model on residuals of the propensity score model and their interactions with the effect modifiers, R-learner attains  Neyman orthogonality \citep{neyman1959optimal} against the two nuisance models,  leading to the same estimation rate as if nuisance models were known  \citep{neyman1959optimal,chernozhukov2018double,nie2021quasioracle}.
Neyman orthogonality was also leveraged in  \cite{kennedy2023towards} to resolve the issue of bias, or equivalently, slow convergence, induced to the target HTE parameter estimation from estimating nuisance models. In general, if a score or loss function satisfies Neyman orthogonality, the estimation error of the nuisance models should not affect the limiting distribution of the estimator for the target parameter \citep{newey1990semiparametric, newey1994asymptotic, bickel1993efficient, van2000asymptotic, chernozhukov2018double, foster2023orthogonal}.

 In healthcare applications, time-to-event outcomes subject to right censoring are often of interest. \citet{xu2023treatment}
 provided a comprehensive tutorial and review that illustrates how popular metalearners (T-learners, S-learners and X-learners) can be extended and used for right censored time-to-event data, with a focus on randomized controlled trials (RCTs). Likewise for time-to-event outcomes, various machine learning based methods have been deployed to the underlying regression problems for HTE estimation
\citep{zhu2020targeted,cui2023estimating}.
However, few existing works provided theoretical guarantees on the estimation rate of the HTE \citep{xu2024estimating,cui2023estimating}. Moreover,
to our best knowledge, the R-learner has not been developed for time-to-event outcomes.
As has been known \citep{hou2023treatment}, the dependence between the at-risk process and the treatment assignment poses a challenge to obtaining orthogonality under certain survival models.

In this work, we propose a new martingale R-learner framework to estimate time-varying HTE for time-to-event outcomes subject to right censoring.
Like its prototype, our martingale R-learner identifies a nonparametric HTE through two nuisance models to be flexibly estimated through machine learning methods:
1) a marginal survival model (over the treatment assignment) for the outcome conditioning on the confounders, and 2) the risk-set propensity score for treatment  that varies along time.
Modeling the risk-set propensity score bypasses the dependence between the treatment and the at-risk process involved in the martingale residual \citep{jiang2017doubly}.
We constructe the martingale R-learner functional score through a decomposition of the martingale residual of the marginal survival model.
Due to the unique semiparametric structure of martingale residuals under survival models,
standard computational and theoretical formulation for metalearners no longer apply.
To develop and analyze the martingale R-learner, we proposed a novel functional score framework that identifies the nonparametric HTE through an estimating equation style approach with  infinite dimensional nuisance models.
We developed a pseudo-data computational scheme to  solve the HTE from the functional score, eventually arriving at a penalized least-squares.
Based on our theoretical groundwork for the general functional score methods, we established that the HTE estimated by our martingale R-learner is insensitive to estimation errors of nuisance models %as the result of Neyman orthogonality
and can achieve the estimation rate as the oracle who knows the nuisance models.

We highlight our major contributions:
\begin{itemize}
    \item We extend the R-learner to time-varying HTE on time-to-event outcomes by developing the martingale R-learner;
    \item We generalize the loss function based metalearner framework to estimating equation style functional score framework;
    \item We prove that the martingale R-learner attains the standard optimal non-parametric estimation rate for smooth HTE.
\end{itemize}

An outline of this paper is as follows. In Section~\ref{sec:notation}, we introduce the general notations and causal settings of the paper.
In Section~\ref{sec:method}, we introduce the functional score framework with its own system of notations and describe the martingale R-learner.
In Section~\ref{sec:theory}, we establish a general theoretical framework for functional scores and apply it to derive the estimation rate of the martingale R-learner.
In Section~\ref{sec:simulation}, we present the evaluation of the martingale R-learner on simulated data.
In Section~\ref{sec:real_data}, we showcase the application in the motivating HAAS study.
In Section~\ref{sec:discuss}, we conclude the paper with the summary statement and discussions on potential extensions.

\csection{Heterogeneous treatment effect} \label{sec:notation}

For subject indexed by $i \in \{1,\dots,n\}$,
let $\bZ_i \in \R^p$ be the p-dimensional vector of baseline covariates collected before the treatment initiation and $\Trt_i$ be the treatment assignment at  time $t=0$. Here we focus on a binary treatment $\Trt_i \in \{0, 1\}$ for our motivating aging study, although our framework applies to bounded discrete or continuous numeric treatments. The outcome of interest is the time-to-event variable $T_i > 0$ subject to the right censoring at time $C_i \in [0,\tmax]$ up to a maximal follow-up time $\tmax$. The observed right-censored outcome is $(X_i,\delta_i)$ with the observation time $X_i = \min\{T_i,C_i\}$ and event status $\delta_i = \ind\{T_i \le C_i\}$. We denote the observed data for each subject as $\datcmb_i = (\bZ_i, \Trt_i, X_i, \delta_i)$ in the sample space $\datspace$, which are assumed to be independent among subjects and follow an identical distribution.
Following the counting process notation \citep{andersen1982cox}, we define the event process $N_i(t) = \delta_i\ind\{X_i\le t\}$ and the at-risk process $Y_i(t) = \ind\{X_i \ge t\}$.

%\csubsection*{Causal inference setting}

To define the HTE, we introduce  $T_i^{(\trt)}$ and $C_i^{(\trt)}$ as the potential time-to-event and censoring time if the treatment were assigned as $\Trt_i = \trt$.
We quantify the HTE using the structural cumulative failure time model (SCFTM):
%defined with the potential outcomes,
\begin{equation}\label{def:SCFTM}
\frac{\P(T_i^{(\trt)} \ge t \mid \bZ_i)}{\P(T_i^{(0)} \ge t \mid \bZ_i)}
= \exp\left\{-  \cumHTE(t,\bZ_i)\cdot \trt\right\},
\end{equation}
which is %the  point exposure version
a special case of the structural nested cumulative failure time models \citep{picciotto2012structural,vansteelandt2014structural}. By imposing the SCFTM \eqref{def:SCFTM}, we characterized the HTE as the log risk ratio  $\cumHTE(t,\bz)$ against the reference or control exposure level $\trt=0$,
and allow it to  vary across both the covariate value $\bz$ and  time $t$.
Like all structural nested models
the definition of the HTE through the structural model~\eqref{def:SCFTM} does not require modeling the distribution of $T_i^{(0)}$.
%the potential outcome under the reference/control exposure $\trt=0$.

Define the survival function $\Surv(t;\bZ_i,\trt) = \P(T_i^{(\trt)} \ge t \mid \bZ_i)$, the cumulative hazard function
    $\Lambda(t;\bZ_i,\trt) = -\log\left\{\Surv(t;\bZ_i,\trt)\right\}$,
    the hazard function     $\lambda(t;\bZ_i, \trt) = \partial  \Lambda(t;\bZ_i,\trt) / \partial t$,
    and the derivative of the log risk ratio
    $\HTE(t;\bZ_i) = \partial  \cumHTE(t;\bZ_i) / \partial t$.
%\end{equation*}
Under model~\eqref{def:SCFTM}
we can alternatively characterize the HTE by
$\HTE(t,\bZ_i)$ as in the followng:
\begin{equation}\label{def:HTE}
   \lambda(t;\bZ_i, \trt)  =
  \lambda(t;\bZ_i, 0) + \trt \cdot \HTE(t,\bZ_i).
\end{equation}

In order to identify as well as make inference about the HTE $\cumHTE(t,\bZ_i)$, or equivalently $\HTE(t,\bZ_i)$, we require the following assumptions, where a.s.~stands for almost surely.
\begin{assumption}[Causal settings]\label{assume:causal}
For  absolute constants $\rate[1], \rate[2] \in (0,1)$, we assume
%that the potential variables and observed data satisfy
\begin{enumerate}[label = (\roman*), ref = \ref{assume:causal}-(\roman*)]
    \item \label{assume:SUTVA} \textbf{Stable Unit Treatment Value Assumption (SUTVA)}. There are no multiple versions of treatment and there is no interaction between units.
    \item \label{assume:cons} \textbf{Consistency}. $T_i = T_i^{(\Trt_i)}$ and $C_i = C_i^{(\Trt_i)}$.
    \item \label{assume:unconf} \textbf{Unconfoundedness}. $\{T_i^{(\trt)}, C_i^{(\trt)}\} \indep \Trt_i \mid \bZ_i$.
    \item \label{assume:pos} \textbf{Strict positivity of treatment}.
    $\inf_{t \in [0,\tmax]}\Var(\Trt_i \mid \bZ_i, X_i \ge t) > \rate[1]$ a.s.~$\bZ_i$.
    \item \label{assume:cen} \textbf{Non-informative censoring}. $C^{(\trt)}_i \indep T^{(\trt)}_i \mid \bZ_i$.
    \item \label{assume:atrisk} \textbf{Strict positivity of at-risk}. $\E\{Y_i(\tmax)\mid\bZ_i\} > \rate[2]$ a.s.~$\bZ_i$.
\end{enumerate}

\end{assumption}

Assumption~\ref{assume:SUTVA} SUTVA and Assumption~\ref{assume:cons} consistency are fundamental prerequisites  for most causal inference methodology \citep{rubin1980randomization, vanderweele2009further}.
In the presence of confounding in observational studies, Assumption~\ref{assume:unconf} unconfoundedness requires that the covariate $\bZ_i$ adequately represent the pre-treatment variables affecting both the treatment assignment and the outcome so that the confounding can be properly accounted for \citep{rosenbaum1983central}.
Assumption~\ref{assume:pos} requires that the treatment has sufficient variability in   risk sets at all time $t$ and all possible covariate $\bz$ values so that HTE can be reflected by contrasting events at different levels of $d$. For the binary treatment $\Trt_i \in \{0,1\}$, Assumption~\ref{assume:pos} is equivalent to the common positivity assumption \citep{rosenbaum1983central} since
$
\Var(\Trt_i \mid \bZ_i, X_i \ge t) =
\P(\Trt_i=1 \mid \bZ_i, X_i \ge t)\P(\Trt_i=0 \mid \bZ_i, X_i \ge t).
$
Assumption~\ref{assume:causal} also reflects the unique role of censoring.
On the one hand, censoring time $C_i^{(\trt)}$ may be affected by treatment assignment, so it is also involved in the Assumptions~\ref{assume:cons} and \ref{assume:unconf}.
On the other hand, censoring is another layer of missing data in addition to the counterfactual outcomes
%on top of the causal inference problem
\citep{ding2018causal,luo2025doubly}, so Assumptions~\ref{assume:cen} and \ref{assume:atrisk} are parallel to Assumptions~\ref{assume:unconf} and \ref{assume:pos} in enabling the estimation of the event time distribution in the presence of right censoring  \citep{cox1972regression,lagakos1979general,lin1994semiparametric}.
All elements of the Assumption~\ref{assume:causal} have been  imposed in the literature for causal inference methods with right-censored time-to-event outcomes \citep{curth2021survite,  hou2023treatment, xu2024estimating}.

As the HTE $\HTE$ in \eqref{def:HTE} is additive for the hazards,  model~\eqref{def:SCFTM} is connect with the additive hazards models \citep{aalen1989linear,martinussen2011estimation}, and our method is intrinsically inspired by the corresponding estimation approaches \citep{lin1994semiparametric,hou2023treatment}.
In particular for model  \eqref{def:HTE} under Assumption~\ref{assume:causal}, we have
%\lily{previously defined for potential outcome}
\begin{equation}\label{def:cond_model}
   \lambda(t;\bZ_i, \Trt_i)  =
  \lambda(t;\bZ_i, 0) + \Trt_i \cdot \HTE(t,\bZ_i).
\end{equation}
As this is a conditional model given $ \Trt_i$,
we define the conditional martingale residual \citep{lin1994semiparametric},
\begin{align}
    \condM_i(t) = & N_i(t) - \int_0^t Y_i(u) \left\{\Trt_i\HTE(u,\bZ_i)+\lambda(u;\bZ_i,0)\right\} du \notag \\
    = & N_i(t) - \int_0^t Y_i(u) \left\{\Trt_i\HTE(u,\bZ_i)du+d\Lambda(u;\bZ_i,0)\right\}.
    \label{def:martingale}
\end{align}
Denote its natural filtration as $\filt_t = \sigma\{\Trt_i, \bZ_i, Y_i(u), N_i(u): u \le t\}$, which represents the $\sigma$-algebra generated by information available at time $t$.

\csection{Orthogonal functional score}\label{sec:method}

 In the following we introduce the Neyman orthogonal functional scores for HTE
as a generalization of Neyman orthogonality for finite dimensional estimators \citep{chernozhukov2018double} and loss function approaches \citep{foster2023orthogonal,kennedy2023towards}.
%traditionally identified through estimating equations in Section~\ref{sec:framework}.
We then demonstrate the derivation of the Martingale R-learner functional score under model~\eqref{def:HTE} according to the idea of R-learner \citep{nie2021quasioracle} and explain its Neyman orthogonality. % in Section~\ref{sec:score}.
%In Section~\ref{sec:est},
We detail the estimation process through a pseudo-data trick that transforms the solution of the functional score into a least-squares optimization problem.

\csubsection{Functional score framework with Neyman orthogonality} \label{sec:framework}

Unlike the original Robinson's estimator that can be formulated directly as regression, leading to Nie and Wager's R-learner using penalized least-squares to estimate the HTE, estimation of $\HTE$ under model~\eqref{def:SCFTM} is motivated by the estimating equations approach that is often employed for various parameters with right-censored time-to-event outcomes \citep{lin1994semiparametric,hou2023treatment}.
%In absence of prior selection of covariates from high-dimensional candidate \citep{huang2013oracle,gaiffas2012high} or knowledge of proper parametric models \citep{kooperberg1995hazard,stone1997polynomial}, the estimation bias and/or uncertainty of such nuisance models can be considerably large.
%Therefore, estimation processes satisfying the Neyman orthogonality \citep{neyman1959optimal,neyman1979c} that can asymptotically eliminate the impact of estimated nuisance models on the causal parameter of interest, HTE $\HTE$ in our problem, are highly desirable.
Existing literature so far has studied Neyman orthogonality for either finite dimensional parameters or infinite dimensional parameters identified by  loss functions \citep{chernozhukov2018double,foster2023orthogonal,kennedy2023towards}.
%Their current frameworks do not cover the
We herein introduce a \emph{functional score} framework that extends the traditional estimating equations for survival data
%for finite dimensional parameters
to nonparametric parameters  and generalize the  Neyman orthogonality concept accordingly, thus  filling a methodological as well as theoretical gap in Neyman orthogonality.

Let $\HTEspace$ and $\nspace$ be two normed infinite dimensional spaces equipped with norms $\HTEnorm{\cdot}$ and $\nnorm{\cdot}$, respectively. The space $\HTEspace$ contains the target parameter of interest, i.e. the HTE in our setting, while the space $\nspace$ contains the nuisance models.
We use $\HTEtrue \in \HTEspace$ and $\NMtrue \in \nspace$ for the ``true parameters''.
We are particularly interested in the dual space for $\HTEspace$, denoted as $\HTEspace^*$ \cite[Definition 6.1.1, page 188]{lang93functional}.
As a generalization of vector-valued scores  in classic parametric statistics,
%which stands for finite dimensional estimating functions,
we define the functional scores
over the space $\HTEspace$. We defer to Section \ref{sec:score} for our martingale R-learner functional score as an example.
\begin{definition}[Functional scores]\label{def:fs}
A functional score $\psi(\HTE,\NM;\datcmb_i)$
is a class of linear functionals in $\HTEspace^*$ indexed by the models $\HTE$ and $\NM$.
For a function $\phi \in \HTEspace$, the element of functional score with models $\HTE$ and $\NM$ maps to the real valued random variable $\psi(\HTE,\NM;\datcmb_i)[\phi]$.  Properly constructed functional score should identify the true parameters $\HTEtrue$ and $\NMtrue$ by satisfying moment equation uniformly in
$\HTEspace$,
$
\E\{\psi(\HTEtrue, \NMtrue; \datcmb_i)[\phi]\} = 0, \, \forall \phi \in \HTEspace.
$
\end{definition}

%Analogous to the matrix valued derivatives of vector valued scores,
We also introduce the derivatives of the functional scores as the functional of the product space \citep{foster2023orthogonal}, indexed by functions $\phi_1,\phi_2,\phi_3$ in spaces $\HTEspace$ or $\nspace$.
\begin{definition}[Derivatives of functional scores]\label{def:Dfs}
For the functional score $\psi$ in Definition~\ref{def:fs}, we define the following derivative functionals
based on the directional derivatives
\begin{align*}
    \Dfun_{\HTE}\psi(\HTE, \NM;\datcmb_i)[\phi_1,\phi_2]
    = & \lim_{\Delta \to 0} \left\{\psi(\HTE+ \Delta \cdot \phi_2, \NM;\datcmb_i)[\phi_1] -
    \psi(\HTE, \NM;\datcmb_i)[\phi_1]\right\}/\Delta; \\
    \Dfun_{\NM}\psi(\HTE, \NM;\datcmb_i)[\phi_1,\phi_2]
    = & \lim_{\Delta \to 0} \left\{\psi(\HTE, \NM +\Delta \cdot \phi_2;\datcmb_i)[\phi_1]-
    \psi(\HTE, \NM;\datcmb_i)[\phi_1]\right\}/\Delta, \\ \,
    \Dfun_{\NM}^2\psi(\HTE, \NM;\datcmb_i)[\phi_1,\phi_2,\phi_3]
    = & \lim_{\Delta \to 0} \left\{\Dfun_{\NM}\psi(\HTE, \NM +\Delta \cdot \phi_3;\datcmb_i)[\phi_1,\phi_2] - \Dfun_{\NM}\psi(\HTE, \NM;\datcmb_i)[\phi_1,\phi_2]\right\}/\Delta.
\end{align*}
\end{definition}
Using the functional score framework specified by Definitions~\ref{def:fs}~and~\ref{def:Dfs}, we can now define
the Neyman orthogonality for functional scores.
\begin{definition}[Neyman orthogonality]\label{def:neyman_fs}
    The functional score $\psi(\HTE, \NM; \datcmb_i)$ with true target parameter $\HTEtrue$ and true nuisance parameter $\NMtrue$
    %as in Definition~\ref{def:fs}
    is Neyman orthogonal if
    $$
    \E\{\Dfun_{\NM}\psi(\HTEtrue, \NMtrue;\datcmb_i)[\HTE-\HTEtrue,\NM - \NMtrue]\} = 0, \; \forall \HTE \in \HTEspace \text{ and } \NM \in \nspace.
    $$
\end{definition}
\begin{remark}\label{remark:neyman_fs}
As
traditional estimating functions can be seen as
%have been used as surrogates for
the gradients of certain (implicit) loss functions %without analytical forms
\citep{wolfson2011eeboost},
%Likewise,
our functional score framework also relates to the loss function approach for infinite dimensional parameters.
Suppose that $\ell(\HTE,\NM;\datcmb_i)$ is a loss function
%$\ell: \HTEspace \times \nspace \times \datspace \mapsto \R$
that identifies  $\HTEtrue$ by
$\HTEtrue = \argmin_{\HTE \in \HTEspace} \E\{\ell(\HTE,\NMtrue;\datcmb_i)\}$.
We denote its directional derivatives similar to Definition~\ref{def:Dfs}.
It can be seen that if $\ell$ is
 locally convex,  i.e.~with $\inf_{\HTEnorm{\phi}=1}\E\{\Dfun_{\HTE}^2\ell(\HTEtrue, \NMtrue;\datcmb_i)[\phi,\phi]\}>0$, then analyzing the one-dimensional sub-problem with optimality of $\HTEtrue$ leads to
$
0 = \argmin_{\Delta \in \R} \E\{\ell(\HTEtrue + \Delta \cdot \phi,\NMtrue;\datcmb_i)\} \text{ and }
\E\{\Dfun_{\HTE}\ell(\HTEtrue,\NMtrue;\datcmb_i)[\phi]\} = 0.
$
This implies that optimization using the loss function $\ell$ can be achieved through the functional score derived from its gradient $\Dfun_{\HTE}\ell$.
The Neyman orthogonality defined for loss function \citep{foster2023orthogonal,kennedy2023towards},
%\lily{double check below}
$$
\E\{\Dfun_{\NM}\Dfun_{\HTE}\ell(\HTEtrue, \NMtrue;\datcmb_i)[\HTE-\HTEtrue,\NM - \NMtrue]\} = 0,  \; \forall \HTE \in \HTEspace \text{ and } \NM \in \nspace,
$$
thus coincides with the Neyman orthogonality defined for the functional score $\psi = \Dfun_{\HTE}\ell$ under Definition~\ref{def:neyman_fs}.
\end{remark}

\csubsection{Functional score for Martingale R-learner} \label{sec:score}

To derive a Neyman orthogonal functional score for HTE $\HTE$ under model~\eqref{def:HTE}, we consider the R-learner framework \citep{robinson1988root,nie2021quasioracle}.
\citet{nie2021quasioracle} considered the conditional mean HTE without right censoring,  and proposed the R-learner that
extended the estimator of \citet{robinson1988root} for the average treatment effect under the partially linear model.
In both works, the key idea  was to decompose the residual from regression of the observed outcome on the covariates, referred to as the
``marginal'' model as it ``integrates'' out the potential outcomes with respect to the treatment assignment.
Let $\ps(\bZ_i) = \P(\Trt_i=1\mid \bZ_i)$ be the propensity score. The residual is then decomposed
into the sum of two uncorrelated components: 1) the residuals of the propensity score $\Trt_i - \ps(\bZ_i)$ multiplied by the treatment effect, and 2) the residual from regression of the outcome on both the treatment and the covariates, i.e.~the
``conditional'' model.
The uncorrelatedness  enables treating the treatment effect as a regression coefficient with the second component above acting as an exogenous error in the regression. In the following we will refer to both the \cite{robinson1988root} and the \cite{nie2021quasioracle} developments as R-learners.
Development of our R-learner falls on the proper construction of the marginal model residuals, the conditional model residuals and the centering of the treatment variable.

We can define
the marginal hazard function of the observed time-to-event outcome $T_i$ given $\bZ_i$:
$\lambda_{m}(t;\bZ_i)
    = \lim_{\Delta \to 0} {\P(T_i < t+\Delta \mid T_i \ge t, \bZ_i)}/{\Delta}$,
    with the corresponding marginal martingale residual
    \begin{equation}\label{def:marg_mart}
        \margM_i(t) = N_i(t) - \int_0^t Y_i(t) d \Lambda_{m}(u;\bZ_i),
    \end{equation}
    where $         \Lambda_{m}(t;\bZ_i) = \int_0^t \lambda_{m}(u;\bZ_i) du$.
The challenge for applying the standard R-learner to HTE $\HTE$ under model~\eqref{def:HTE} %with right censored time-to-event outcomes,
is that the typical martingale residual \eqref{def:martingale} for a conditional model~\eqref{def:cond_model} is often correlated with the residuals of the propensity score variable $\Trt_i - \ps(\bZ_i)$ \citep{kang2018estimation,dukes2019doubly,hou2023treatment}. To decouple the correlation, we consider
%the adjustment to the centering of treatment effect by
the approach of \cite{kang2018estimation} using risk-set propensity score:
\begin{equation}\label{def:risk-ps}
    \rps(t,\bZ_i) = \E\{\Trt_i \mid Y_i(t) = 1, \bZ_i\}.
\end{equation}
An alternative approach by \cite{dukes2019doubly} and \cite{hou2023treatment} involved a multiplicative factor of risk ratio   but  that turned out to be difficult for use here.

\begin{lemma}[R-learner martingale residual]\label{lemma:martingale}
Under the model~\eqref{def:HTE} and Assumption~\ref{assume:causal}, we have the decomposition
of marginal martingale residual \eqref{def:marg_mart},
$$
\margM_i(t) = \int_0^t \HTE(u,\bZ_i)Y_i(u)\{\Trt_i - \rps(u,\bZ_i)\}du + \condM_i(t),
$$
with the conditional martingale residual $\condM_i$ in \eqref{def:martingale}, HTE $\HTE$ in \eqref{def:HTE} and the residuals of the risk-set propensity score $\rps$ in \eqref{def:risk-ps}.
\end{lemma}

The proof of Lemma~\ref{lemma:martingale} (see Section~\ref{app:proof_martingale} in the Supplementary Materials) is based on the following fact:
$$
\lambda_m(t;\bZ_i) = \rps(t,\bZ_i)\HTE(t,\bZ_i) + \lambda(t;\bZ_i, 0).
$$
Lemma~\ref{lemma:martingale} provides the canonical form of the R-learner construction \citep{robinson1988root,nie2021quasioracle}: the left-hand side is the marginal martingale residual; the first term on the right is the integral of a predictable process under filtration $\filt_t$ from the product of the causal parameter $\HTE$ with the residuals of the risk-set propensity score; the second term on the right is the conditional martingale residual.
The increments of the the $\filt_t$--predictable process and $\filt_t$--martingale on the right are uncorrelated according to standard martingale transforms argument \cite[Theorem 5.2.5]{Durrett_2019}.
The involved nuisance models are the risk-set propensity score $\rps$ in \eqref{def:risk-ps} and marginal model cumulative hazard $\Lambda_m$ within the marginal martingale residual $\margM_i$ through \eqref{def:marg_mart}.

Unlike the standard R-learners, we cannot directly regress $\margM_i(t)$ onto the residuals of the risk-set propensity score $\Trt_i - \rps(t,\bZ_i)$ because the $\condM_i(t)$'s are not i.i.d.~and cannot act as regression errors. Instead,
%increments of counting process martingales, as oppose to the martingales themselves, are the effective residuals in survival analyses \citep{lin1994semiparametric}.
applying the functional score framework, we propose to identify the HTE by the following martingale R-learner functional score
\begin{align}
    \fsMRL(\HTE,\rps,\Lambda_m;\datcmb_i)[\phi]
    =& \int_0^{\tmax} \phi(t,\bZ_i) \{\Trt_i-\rps(t,\bZ_i)\}\left\{dN_i(t) - Y_i(t)d\Lambda_m(t;\bZ_i)\right\} \notag \\
    &-\int_0^{\tmax} \phi(t,\bZ_i) \HTE(t,\bZ_i) Y_i(t) \{\Trt_i-\rps(t,\bZ_i)\}^2 dt. \label{def:fs_mrl}
\end{align}
where $\phi(t,\bZ_i) \in \HTEspace$
%$= \{\phi: \HTEnorm{\phi}<\infty\}$
is an arbitrary square integrable function, and
 $\NM = (\rps,\Lambda_m)$  are the nuisance models.

We verified that the martingale R-learner functional score \eqref{def:fs_mrl} can identify the true $\HTEtrue$ and attains the Neyman orthogonality (Definition~\ref{def:neyman_fs}).
\begin{lemma}[Neyman orthogonality]\label{lemma:neyman_mrl}
Under Assumptions~\ref{assume:causal} and  model~\eqref{def:HTE}, the martingale R-learner functional score $\fsMRL$  in \eqref{def:fs_mrl} satisfies
 %   \begin{gather*}
   $\E\left\{\fsMRL(\HTEtrue,\rpstrue,\mLamtrue;\datcmb_i)[\phi]\right\} =$   \\
    $\E\left\{\Dfun_{\rps}\fsMRL(\HTEtrue,\rpstrue,\mLamtrue;\datcmb_i) [\phi,\rps-\rpstrue]\right\} = %0, \\
    \E\left\{\Dfun_{\Lambda_m}\fsMRL(\HTEtrue,\rpstrue,\mLamtrue;\datcmb_i)[\phi,\Lambda_m-\mLamtrue]\right\} = 0$,
%    \end{gather*}
    at the true parameters $\HTEtrue$, $\rpstrue$, $\mLamtrue$ for any
    $\phi \in \HTEspace$ and $(\rps,\Lambda_m) \in \nspace$.
\end{lemma}

The proof of Lemma~\ref{lemma:neyman_mrl} (see Section~\ref{app:proof_neyman} in the Supplementary Materials) anchors on the alternative form of \eqref{def:fs_mrl} as the product of two uncorrelated residuals following Lemma~\ref{lemma:martingale},
$$
\fsMRL(\HTEtrue,\rpstrue,\mLamtrue;\datcmb_i)[\phi]
= \int_0^{\tmax} \phi(t,\bZ_i) \{D_i -\rpstrue(t,\bZ_i)\} d\condM_i(t).
$$
%Many existing literature utilized the same structure to establish the Neyman orthogonality for various causal parameters \citep{chernozhukov2018double,nie2021quasioracle,hou2023treatment}.
Following Lemma~\ref{lemma:neyman_mrl}, we may estimate the risk-set propensity scores \eqref{def:risk-ps} and marginal model~\eqref{def:marg_mart} using flexible machine learning approaches with moderately slow convergence rates and construct a consistent estimator for the HTE with oracle convergence rates.

\csubsection{Implementation}\label{sec:est}

We first discuss the estimation of the two nuisance parameters $\rho$ and $\Lambda_m$ involved in the martingale R-learner \eqref{def:fs_mrl}.
%To reduce the artificial estimation bias from
In order to guarantee compatibility, we propose to construct both nuisance models  from two shared basic models: 1) the (baseline) propensity scores $\ps(\cdot)$, and 2) the conditional survival function $\Surv(t;\bZ_i,\trt) = \P(T_i \ge t\mid \bZ_i, \Trt_i = \trt)$.
%of event time $T_i$ conditioning on covariates $\bZ_i$ within treatment groups derived from
For the risk-set propensity scores $\rho$ in \eqref{def:risk-ps}, under the stronger assumption of noninformative treatment for censoring  given the covariates,
we have
%considered an alternative estimation approach based on the another expression,
\begin{equation}\label{eq:risk-ps-alt}
    \rps(t,\bZ_i) = \P(\Trt_i=1\mid\bZ_i,X_i\ge t)
 = \frac{\ps(\bZ_i)\Surv(t;\bZ_i,1)/\Surv(t;\bZ_i,0)}{1+\ps(\bZ_i)\left\{\Surv(t;\bZ_i,1)/\Surv(t;\bZ_i,0) - 1\right\}};
\end{equation}
we provide the proof in Section~\ref{app:rps} of the Supplementary Materials as well as extension to potential treatment-dependent censoring (Assumption~\ref{assume:cen}).
Likewise for $\Lambda_m$ in
%the marginal model~
\eqref{def:marg_mart}, we derive the alternative expression
\begin{equation}\label{eq:margHaz-alt}
    \Lambda_m(t ; \bZ_i) = -\log \left[\ps(\bZ_i)\Surv(t;\bZ_i,1)
    + \{1-\ps(\bZ_i)\}\Surv(t;\bZ_i,0)\right].
\end{equation}
% any statistical regression or machine learning methods for right-censored outcomes that produces the estimated conditional survival function or cumulative hazard function given covariates \citep{ishwaran2008random,kooperberg1995hazard} can be deployed.
% Using the outcomes $(\delta_i,X_i)$ and the covariates $\bZ_i$,
% %without treatment variable $\Trt_i$,
% we directly obtain the estimated $\hat{\Lambda}_m(t;\bZ_i)$, integrated from estimated hazard function $\hat{\lambda}_m(t;\bZ_i)$, or derive from the estimated survival function $\hat{\Surv}_{m}(t;\bZ_i)$ through $\hat{\Lambda}_m(t;\bZ_i) = -\log\left\{\hat{\Surv}_{m}(t;\bZ_i)\right\}$.

Under a cross-fitting scheme, we split the data into $K$ folds of roughly equal sizes $n_k$ whose indices are $\fold_{k}$, $k=1,\dots,K$. Denote the out-of-fold indices as $\fold_k^c = \{1,\dots,n\}\setminus\fold_k$
and the out-fold-data as $\datout = \{\datcmb_i: i\in \fold_k^c\}$.
We estimate the basic models $\psk$ and $\Survk$ using the out-of-fold data $\datout$
%defined in Section \ref{sec:notation} \hj{first paragraph} \lily{where}
to construct the out-of-fold estimators $\rpsk$ and $\mLamk$ according to \eqref{eq:risk-ps-alt} and \eqref{eq:margHaz-alt} with $\hat{\pi}$ and $\hat{\Surv}$ obtained from flexible estimation.

As the infinite dimensional spaces $\HTEspace$ and $\nspace$ can be overwhelmingly large, we may consider much smaller subsets for feasible estimators.
For the HTE we consider the $q$-dimensional approximation  $\HTEspace_q \subseteq \HTEspace$
following the classic polynomial splines in  nonparametric estimation \citep{stone1982optimal,stone1997polynomial},
  by
\begin{equation}\label{def:basis}
    \HTE(t,\bZ_i) \approx \Vcoef^\top\Vbasis(t,\bZ_i)
    = \sum_{j=1}^q \coef_j \basis_j(t,\bZ_i),
\end{equation}
where $\basis_j$ are basis functions of space $\HTEspace$ such as the natural splines or B-splines.
In our functional score framework, the basis functions not only approximates the HTE $\HTE$ through \eqref{def:basis} but also guide the approximation of the arbitrary $\phi$ for the infinite dimensional equation in Definition~\ref{def:fs}.
We propose to estimate the HTE by solving $\hat{\Vcoef}$ by utilizing the $q$-dimensional cross-fitted estimating equation
%constructed with in-fold-$k$ data  and out-of-fold estimators $\rpsk$ and $\mLamk$,
%\lilyemph{$ \basis_j \neq \phi$ though}
\begin{equation}\label{eq:EE_basis}
    \frac{1}{n}\sum_{k=1}^K\sum_{i\in\fold_k} \fsMRL\left(\Vcoef^\top\Vbasis, \rpsk,\mLamk;\datcmb_i\right)[\basis_j] = 0, \, j=1,\dots,q.
\end{equation}
We considered two strategies, directly solving equation~\eqref{eq:EE_basis} or approximately solving the equation with regularization of $\Vcoef$ (see Remark~\ref{remark:algorithm} and Section~\ref{app:algorithm} of the Supplementary Materials). Plugging the solution $\hat{\Vcoef}$ into the basis approximation~\eqref{def:basis}, we obtain the estimated HTE, $\hat{\HTE}(t,\bz)=\hat{\Vcoef}^\top\Vbasis(t,\bz)$.
This time-varying  HTE can distinguish short-term versus long-term effects, informing the \emph{infinitesimal effect} of treatment for subject with covariate $\bz$ at time $t$.
%When a univariate metric is of interest
On the other hand, for cumulative risks an estimator for the risk ratio in \eqref{def:SCFTM} can be derived from
$\hat{\cumHTE}(t,\bz) = \int_0^t\hat{\HTE}(t,\bz)dt$.

\begin{remark}\label{remark:algorithm}
Notice that %the martingale R-learner functional score $\fsMRL$ \eqref{def:fs_mrl} is linear in the HTE $\HTE$.
%With the linear basis approximation \eqref{def:basis},
%the estimating equation \eqref{eq:EE_basis} is a $q$-dimensional system of equations with closed form solutions.
%However,
the evaluation of the integrals  \eqref{def:fs_mrl} involving the estimated $\rpsk$ and $\mLamk$ usually has no immediate analytical solutions.
In Section~\ref{app:algorithm} of the Supplementary Materials
we show that a pseudo data based algorithm  can be solved using weighted least squares  with a memory friendly divide-and-conquer option for large datasets. The pseudo least squares also enable the convenient extension to regularized splines by adding the penalty term $\eta \pennorm{\Vcoef}$ with the norm of choice $\pennorm{\cdot}$ \citep{eilers1996flexible}.
As cross-validation or its computationally efficient alternatives have been recommended to determine the choices of splines
\citep{oSullivan1988nonparametric,ruppert2002selecting},
we proposed to use the same pseudo least squares to select dimension $q$ and basis functions $\basis_j$ for splines, as well as the penalty factor $\eta$ for regularized splines.
\end{remark}

The theoretical choice of dimension $q$ and basis functions $\basis_j$ has been thoroughly studied for classic nonparametric regressions \citep{stone1982optimal,stone1997polynomial}. In Section~\ref{sec:theory} below, we establish a theory guided oracle choice of basis dimension $q$ for estimation from \eqref{eq:EE_basis} in order to achieve the standard estimation rate of polynomial splines as if the nuisance models were known.
The theory also applies to its regularized splines with penalty factor $\eta$ decaying in appropriate rate.

\csection{Asymptotic theory}\label{sec:theory}

Like in the R-learner \citep{nie2021quasioracle} or general
orthogonal learning \citep{foster2023orthogonal},
the centerpiece of our theoretical analysis is the oracle property.
With the oracle property,
the estimation error of nuisance models
does not impact the convergence rate of the estimated target parameter,
i.e.~using the estimated nuisance models
is as good as  knowing the true nuisance models.
This leads to the estimation for HTE at the rate previously established as optimal for nonparametric regressions.
%(see Remark~\ref{remark:optimality}).
In the following we first establish a general theory in Section \ref{sec:theory_general} for the oracle property of Neyman orthogonal functional scores
under broad regularity conditions.
We then apply the general theory to the Martingale R-learner with specified assumptions to derive the
estimation error rate in Section \ref{sec:theory_mrl}.
Throughout our theory, we assume the folds $k=1,\dots,K$ are exchangeable with shared theoretical properties and hence suppress the fold notation for cross-fitted models.

\csubsection{Theory for general functional score}\label{sec:theory_general}

We first establish a parallel result to Theorem~1 of \cite{foster2023orthogonal} for functional score.
Using the notations and definitions  in Section \ref{sec:framework},
we introduce the following regularity conditions sufficient for oracle property. Recall that $\HTEtrue \in \HTEspace$ and $\NMtrue \in \nspace$ is the true
target parameter and true nuisance parameter,  and the subspaces $\hat{\HTEspace} \subseteq \HTEspace$ and $\hat{\nspace} \subseteq \nspace$ contain possible estimators for $\HTE$ and $\NM$, respectively.

%When the functional score $\psi$ is derived from the derivative of a loss function with respect to the target parameter,
The regularity conditions in Assumption \ref{assume:general} below for the functional score $\psi$
are parallel to those
required for orthogonal learning through
loss functions by \citet{foster2023orthogonal}.
%which covers many existing semi-parametric estimation methods utilizing Neyman orthogonality \cite[among others]{chernozhukov2018double,nie2021quasioracle,kennedy2023towards}.
% Our Assumption \ref{assume:neyman} is the extension of the ``orthogonal loss'' (their Assumption 1) to the orthogonal functional scores.
% Substituting the identification of true parameters by the ``first-order optimality'' (their Assumption 2) for the loss function,
% our Assumption \ref{assume:identify} is
% an estimating equation style moment condition.
% \lily{I would move (i)(ii) out of Assumption 2 as set up/premise; it is unusual for Foster to list them as assumptions and we don't have to do the same. Assumption 2 otherwise looks too bulky and unnatural. The above paragraph can then be merged}
In case that the spaces of estimators fail to cover the true parameters or models, we leverage the cone set
\begin{equation}
    \starset{\HTEtrue, \hat{\HTEspace}} = \left\{\Delta (\HTE - \HTEtrue): \HTE \in \hat{\HTEspace}, \Delta \in [0,1] \right\}.
    \label{def:star}
\end{equation}
and the best approximation $\HTEq = \argmin_{\HTE \in \HTEspace_q} \HTEnorm{\HTE - \HTEtrue}$.

%To facilitate the cross-fitting to be used later,
We use $\datcmb_{\inew} = (\bZ_{\inew}, \Trt_{\inew}, X_{\inew}, \delta_{\inew})$ for the data for a new observation following the same distribution yet independent of the observed data,
and denote the expectation with respect to $\datcmb_{\inew}$ conditioning on the observed data as $\E_{\inew}\{f(\datcmb_{\inew};\datcmb_1,\dots,\datcmb_n)\}$.

\begin{assumption}[Regularity for oracle property]\label{assume:general}
For  absolute constants $\const{1}, \const{2},\const{3},\const{4} \in (0,\infty)$ and $\rate[3] \in [0,1)$,
%and candidate estimated models,
%        $\forall \HTE \in \hat{\HTEspace}$ and $\forall \NM \in \hat{\nspace}$,
%        and functions in the cone~set~\eqref{def:star}, $\forall \phi_1 \in \starset{\HTEtrue,\hat{\HTEspace}}$ and  $\forall \phi_2 \in \starset{\NMtrue,\hat{\nspace}}$,
        we assume that the general score $\psi(\HTE,\NM; \datcmb_i)[\phi]$
satisfies the following.
%\lilyemph{enumerating is probably taking up too much space} \hj{adjust indent? Will take more space if listed as separate assumptions.}
    \begin{enumerate}[label = (\roman*), ref = \ref{assume:general}-(\roman*)]
%         \item \label{assume:neyman} \textbf{Neyman orthogonality}.    The score $\psi$ is Neyman orthogonal (Definition~\ref{def:neyman_fs}).
%         \item \label{assume:identify} \textbf{Unbiasedness}. The true  target parameter $\HTEtrue$ and nuisance model $\NMtrue$ can be identified with
%         functions in the cone~set~\eqref{def:star},
% $$
% \E\{\psi(\HTEtrue,\NMtrue; \datcmb_i)[\phi]\} = 0, \; \forall \phi \in \starset{\HTEtrue,\hat{\HTEspace}}.
% $$
        \item \label{assume:smooth} \textbf{Smoothness}. The  derivatives (Definition~\ref{def:Dfs})
        are uniformly continuous:
\begin{gather*}
 \sup_{\phi_1 \in \starset{\HTEtrue,\hat{\HTEspace}}}    \frac{\E\left\{ \Dfun_{\HTE} \psi(\HTEtrue+\phi_1,\NMtrue;\datcmb_i) \left[\phi_1,
 \phi_1\right] \right\}}{\HTEnorm{\phi_1}^2} \leq \const{1},\\
\sup_{\phi_1 \in \starset{\HTEtrue,\hat{\HTE}},
\phi_2,\phi_3\in\starset{\NMtrue,\nspace}}
\frac{\left|\E\left\{
\Dfun^2_{\NM}\psi\left(\HTEtrue, \NMtrue + \phi_2\right)\left[\phi_1, \phi_3, \phi_3\right]\right\}\right|}{\HTEnorm{\phi_1}^{1-\rate[3]} \nnorm{\phi_3}^2}
\leq  \const{2}.
\end{gather*}

       \item \label{assume:convex} \textbf{Invertibility}. The  first-order derivatives with respect to the target parameter (Definition~\ref{def:Dfs})
        is bounded away from zero
        for $\forall \phi_1,\phi_2 \in \starset{\HTEtrue,\hat{\HTEspace}},
\phi_3 \in \starset{\NMtrue,\nspace}$:
\begin{equation*}
\E\left\{ \Dfun_{\HTE} \psi(\HTEtrue+\phi_1,\NMtrue + \phi_3;\datcmb_i) \left[\phi_2,
\phi_2\right] \right\}\ge \const{3} \cdot \HTEnorm{\phi_2}^2 - \const{4}\nnorm{\phi_3}^{4/(1+\rate[3])}. % \lilyx{>0??}.
\end{equation*}
%\hj{$>0$ not needed. The first term is positive and second term will be move to the other side of inequality in proof}

\item \label{assume:rate_gen}
\textbf{Convergence of estimators}. For an independent testing data $\datcmb_{\inew}$, the  target parameter estimator $\hat{\HTE}$ and the cross-fitted nuisance model $\NMh$ satisfy
$$
\E_{\inew}\left\{\psi\left(\hat{\HTE},\NMh;\datcmb_{\inew}\right)[\hat{\HTE}-\HTEtrue]\right\}+\nnorm{\NMh-\NMtrue} = \smop{1}.
$$
% \hj{Changed $\NMk$ in the expectation to $\NMtrue$. Should still true under Assumptions \ref{assume:neyman} and \ref{assume:smooth}.}
\end{enumerate}
\end{assumption}
Our Assumptions \ref{assume:smooth} and \ref{assume:convex}, like ``higher-order smoothness'' and ``strong convexity'' in \citet[their Assumptions 3 \& 4]{foster2023orthogonal}, controls the local geometry of the functional score.
Assumption \ref{assume:smooth} imposes local smoothness of the functional score around the true parameters, leading to the local asymptotic linearity of the function score with respect to the HTE. Assumption \ref{assume:convex} guarantees the uniqueness of local solution.
Assumption~\ref{assume:rate_gen} requires that the estimated nuisance models are consistent and the target estimator asymptotically solves the functional score equation in Definition~\ref{def:fs}.
This replaces the ``excess risks'' under the loss function formulation and should be expected for estimators obtained from solving equations like \eqref{eq:EE_basis}, which is an example of the following general functional equation,
\begin{equation}\label{eq:EE_general}
    \frac{1}{n}\sum_{k=1}^K\sum_{i\in\fold_k} \psi(\hat{\HTE}, \NMk; \datcmb_i)[\phi] = 0, \,
    \forall \phi \in \hat{\HTEspace}.
\end{equation}
While the convergence in Assumption~\ref{assume:rate_gen} were not explicitly declared by \citet{foster2023orthogonal},
their corresponding rates for nuisance model and excess risks must converge to zero to establish the consistency of target estimator.
For any loss function satisfying the assumptions for Theorem 1 of \citet{foster2023orthogonal},
the functional score derived from its derivative, as discussed in Remark \ref{remark:neyman_fs}, would satisfy Assumption~\ref{assume:general}.
The reverse problem for a given functional score is substantially harder, and the anti-derivative loss function might not exist.

We state the theory for general Neyman orthogonal functional scores.
\begin{theorem}[General functional score]\label{thm:general}
Let $\hat{\phi} = \left(\hat{\HTE}-\HTEtrue\right)/\HTEnorm{\hat{\HTE}-\HTEtrue}$ be the standardized estimation error.
For the Neyman orthogonal functional score $\psi$ (Definitions~\ref{def:fs}~and~\ref{def:neyman_fs}) and estimators $\{\hat{\HTE}, \NMk\}$ satisfying
Assumption~\ref{assume:general},
the estimation error rate is given by
$$
\HTEnorm{\hat{\HTE}-\HTEtrue}
= \bgOp{
\E_{\inew}\left\{\psi\left(\hat{\HTE},\NMh;\datcmb_{\inew} \right)\left[\hat{\phi}\right] \right\}+\nnorm{\NMh-\NMtrue}^{\frac{2}{1+\rate[3]}}} = \smop{1}.
$$
% \hj{Changed $\NMk$ in the expectation to $\NMtrue$. Should still true under Assumptions~\ref{assume:neyman} and~\ref{assume:smooth}.}
\end{theorem}

%Besides the consistency of suitable functional score estimators under Assumption~\ref{assume:rate_gen}, the conclusion of
Theorem \ref{thm:general}
provides the estimation rate that reflects
the potential oracle property.
The term $\E\{\psi(\HTEtrue,\NM;\datcmb_i)[\HTE-\HTEtrue]\}$ represents the standard generalization error of the equation \eqref{eq:EE_general} solved %over observed sample and
within the subspace $\hat{\HTEspace}$,
which typically determines the estimation rates in classical non-parametric estimation \citep{newey1997convergence}.
For $\rate[3] \in [0,1)$ as specified in Assumption~\ref{assume:general}, the order for estimation error of the nuisance model is higher than the first order if $2/(1+\rate[3])>1$.
That means we may still have the oracle property
$$
\HTEnorm{\hat{\HTE}-\HTEtrue}
= \bgOp{
\E_{\inew}\left\{\psi\left(\hat{\HTE},\NMh;\datcmb_{\inew} \right)[\hat{\HTE}-\HTEtrue] \right\}},
$$
i.e. the estimation rate is not affected by the slower estimation rate of nuisance model,  as long as its $2/(1+\rate[3])$-power is faster
$$
\nnorm{\NMh-\NMtrue}^{\frac{2}{1+\rate[3]}} \ll
\E_{\inew}\left\{\psi\left(\hat{\HTE},\NMh;\datcmb_{\inew} \right)[\hat{\HTE}-\HTEtrue] \right\}
\le \nnorm{\NMh-\NMtrue}.
$$
Among literatures on estimating finite dimensional parameters at $n^{-1/2}$ rate, the requirement for nuisance model estimation is typically $n^{-1/4}$ root mean square error (RMSE) rate, the square root of target $n^{-1/2}$ rate \citep{chernozhukov2018double}.
Although our minimax estimation rate for nonparametric HTE $\HTE$ is slower than $n^{-1/2}$ \citep{stone1982optimal}, our required estimation rate for nuisance model is also the square root of target rate for HTE when $\rate[3]=0$ and slower than $n^{-1/4}$ rate accordingly.

The detailed proof is provided in Section~\ref{app:proof_general} of the Supplementary Materials.
Our proof is anchored on the construction of the pseudo regret (excess risk) function,
\begin{equation}\label{def:pseudo_regret}
\regret{\HTE,\NM} =
    \E_{\inew}\{\psi(\HTE,\NM;\datcmb_{\inew})[\HTE-\HTEtrue]\}
    - \E_{\inew}\{\psi(\HTEtrue,\NM;\datcmb_{\inew})[\HTE-\HTEtrue]\}.
\end{equation}
Note that~\eqref{def:pseudo_regret} is the contrast of the functional scores with parameter $\HTE$ against the that with true $\HTEtrue$ at the estimation error $\HTE-\HTEtrue$.
Its construction is inspired by the symmetrized Bregman divergence
% , the inter-product between estimation error in coefficients and contrasts in the gradient of the loss function,
used in the theory for high-dimensional regressions as an alternative to the Kullback-Leibler (K-L) divergence \citep{huang2012estimation}, where the
estimator is  typically identified by solving the quasi-moment Karush-Kuhn-Tucker (KKT) conditions.
We showed in our proof that the pseudo regret~\eqref{def:pseudo_regret} can take the same role as the excess risks in \citet{foster2023orthogonal} for theoretical analysis of the functional scores.

\csubsection{Theory for Martingale R-learner}\label{sec:theory_mrl}

We now apply Theorem \ref{thm:general} to establish the estimation error rate of HTE from  the Martingale R-learner. Specifically for $\phi(t,\bZ_i)$ in the space $\Hcal$ of square integrable functions that contains the HTE under model \eqref{def:HTE},
 denote the inner product and its induced $L_2$ norm, the uniform norm, %another $L_2$ norm \citep{kooperberg1995l2}
and the dual norms for measures as
\begin{gather}
\HTEprod{\phi_1}{\phi_2} = \int_0^{\tmax} \E\{\phi_1(t,\bZ_i)\phi_2(t,\bZ_i)\}dt, \;
\MSEnorm{\phi} = \sqrt{\HTEprod{\phi}{\phi}}, \;
\Unifnorm{\phi} = \sup_{t\in [0,\tmax]} \Unifnorm{\phi(t,\cdot)}, \notag \\
\HTEdual{\phi_1 d\mu} = \sup_{\phi_2 \in \HTEspace, \HTEnorm{\phi_2}=1} \E\left\{\int_0^{\tmax} \phi_1(t,\bZ_i)\phi_2(t,\bZ_i)d \mu(t;\bZ_i) \right\}, \;
\HTEdual{\phi_1} = \HTEdual{\phi_1 d t}.
% \notag \\
% \HTEqdual{\phi_1 d\mu} = \sup_{\phi_2 \in \HTEspace_q, \HTEnorm{\phi_2}=1} \E\left\{\int_0^{\tmax} \phi_1(t,\bZ_i)\phi_2(t,\bZ_i)d \mu(t;\bZ_i) \right\},  \;
% \HTEqdual{\phi_1} = \HTEqdual{\phi_1 d t}.
    \label{def:HTEnorms}
\end{gather}
Under mild regularity conditions, the norms $\MSEnorm{\cdot}$,  $\HTEdual{\cdot}$ and another $L_2$ norm \citep{kooperberg1995l2} are equivalent (see the Section~\ref{app:norms} of the Supplementary Materials).

The following regularity conditions along with Assumption~\ref{assume:causal} are shown to be sufficient for
Assumption~\ref{assume:general}, in Section~\ref{app:proof_MRL} of the Supplementary Materials.

\begin{assumption}[Regularity for Martingale R-learner]\label{assume:MRL}
For  absolute constants $\const{5},\const{6},\const{7} \in [1,+\infty)$ and a sequence $\basU \ge 1$ indexed by the dimension $q$ of basis in~\eqref{def:basis},
%and norms $\MSEnorm{\cdot}$, $\Unifnorm{\cdot}$ defined in \eqref{def:HTEnorms},
we assume the following.
\begin{enumerate}[label = (\roman*), ref = \ref{assume:MRL}-(\roman*)]
   \item \label{assume:cumhaz-TV} \textbf{Bounded marginal hazards}. The measure induced by the estimated marginal cumulative hazards $\mLamh$ admits Radon-Nikodym representation with respect to a mixed Lebesgue-counting measure whose continuous and jumping components are characterized by $\mlamht$ and $\mlamhn$, respectively,
   $$
   \mLamh(t;\bZ_i) = \int_0^t \mlamht(u;\bZ_i) du
   + \frac{\mlamhn(u;\bZ_i)}{\Nlamh(\tmax)} d\Nlamh(u), \;
   \Unifnorm{\mlamht}+\Unifnorm{\mlamhn} \le \const{5},
   $$
   where $\Nlamh$ is a counting measure over a finite set determined by training data of $\mLamh$.
%    \textbf{Bounded marginal hazards}. The estimated marginal cumulative hazards $\mLamk$ have bounded total variation almost surely,
% $$
% \bigvee_{t=0}^{\tmax} \mLamk(t;\bZ_i) =
%  \sup_{\substack{K=1,2,3, \dots \\ 0=t_0<\dots<t_K=\tmax}} \sum_{k=1}^K |\mLamk(t_k;\bZ_i)-\mLamk(t_{k-1};\bZ_i)|\le \const{5}.
% $$
    \item \label{assume:donsker} \textbf{Donsker class of models}. The bounded subset for HTE,
    and the candidate nuisance models
    $\hat{\nspace}_{\NM}$ belong to a Donsker class $\donsker$ of functions $f(\bz)$ of covariates $\bZ_i$: %\lilyx{such that}
    \begin{gather*}
     \{\HTEtrue/\const{6}, \hat{\HTE}/\const{6}\} \subseteq   \HTEunit = \left\{\phi(t,\bz)\in \HTEspace: \MSEnorm{\phi} \le 1, \, \Unifnorm{\phi} \le 1, t \in [0,\tmax] \right\} \subseteq \donsker, \\
    \hat{\nspace}_{\NM} = \{\rpsh(t,\bz), \mlamht(t;\bz), \mlamhn(t;\bz): t \in [0,\tmax]\} \subseteq \donsker.
    \end{gather*}
    %are function of $\bz$ indexed by  $t$ and $g$.

   \item \label{assume:norm}
   \textbf{Uniform norm in subspace}.
   Over finite dimensional basis expansion subspace $\Hcal_q$ spanned by the basis in~\eqref{def:basis},
   the uniform norm has bound
    $
    \sup_{\phi \in \Hcal_q, \MSEnorm{\phi}=1} \Unifnorm{\phi} \le \basU
    $.

    \item \label{assume:rate_MRL} \textbf{Consistent nuisance models}.
    The nuisance models are consistent in suitable norms,
    \begin{gather*}
    \left(\rpsnorm{\rpsh-\rpstrue} + \Lamnorm{\mLamh-\mLamtrue}\right)
    = \smop{1}, \; \HTEdual{\left(\rpsh-\rpstrue\right)^2} \le \const{7} \rpsnorm{\rpsh-\rpstrue}^2, \\
    \HTEdual{\left(\rpsh  -\rpstrue\right) d\left(\mLamh-\mLamtrue\right)} \le
    \const{7} \rpsnorm{\rpsh-\rpstrue}\Lamnorm{\mLamh-\mLamtrue}.
    \end{gather*}
\end{enumerate}
\end{assumption}

Assumption~\ref{assume:cumhaz-TV} is applicable for various flexible estimation methods of marginal survival models with continuous or discrete $\mLamh$ \citep{ishwaran2008random,kooperberg1995hazard},  as long as their total variation is bounded, which is guaranteed for $\mLamh(\cdot)$ monotone in time.
As the basis approximation \eqref{def:basis} usually
require the HTE to belong to a smooth class,
the Donsker property for such a class in Assumption~\ref{assume:donsker} is typically well established \cite[Example 2.10.25]{vandervaart1996weak}.
Likewise, the Donsker property for the class of estimated nuisance models $\rpsh$, $\mlamht$, and $\mlamhn$ in Assumption~\ref{assume:donsker} should be satisfied with classical nonparametric estimation methods that produce smooth estimated models.
While Donsker class admits uniform weak convergence of empirical processes at $\sqrt{n}$ rate \citep{vandervaart1996weak}, nonparametric estimation of models from infinite dimensional Donsker class is usually slower than the parametric rates \citep{stone1982optimal}.
For splines, increasing dimensionality usually creates basis with higher resolution that concentrates in a smaller area, which creates larger ratio between uniform and $L_2$ norms for these basis functions, e.g. piecewise polynomial splines satisfy Assumption~\ref{assume:norm} with $\basU = O(\sqrt{q})$   \citep{donald1994series}.
Assumption~\ref{assume:rate_MRL} is the consistency of nuisance models in Assumption~\ref{assume:rate_gen} for the Martingale R-learner. The required norms are usually guaranteed by uniform convergence of $\rpsh$ and $\mLamh$ but may also be satisfied by $L_2$ convergence under regression models with light-tailed covariates \citep{hou2023treatment}.

We now state the theory for the Martingale R-learner.
%Let  $\HTEq$ , and the norms be defined in \eqref{def:HTEnorms}.
To cover both the standard and regularized splines, we characterize the estimated HTE by its proximity of solving the estimating equation~\eqref{eq:EE_basis},
\begin{equation}\label{def:Snq}
    \hatscore = \sup_{\phi \in \HTEspace_q, \HTEnorm{\phi}=1}\frac{1}{n} \sum_{k=1}^K\sum_{i\in\fold_k}  \fsMRL\left(\hat{\HTE},\rpsk, \mLamk;\datcmb_i \right)[\phi].
\end{equation}
For standard splines solved precisely from equation~\eqref{eq:EE_basis}, we have $\hatscore = 0$ due to linearity of functional score with respect to $\phi$ and linear representation of any $\phi \in \HTEspace_q$ by $\basis_j$. For regularized splines, $\hatscore$ is typically determined by norm $\pennorm{\cdot}$ and penalty factor $\eta$ through analyzing the KKT condition (see Section~\ref{app:algorithm} of the Supplementary Materials).
Recall the best approximation $\HTEq = \argmin_{\HTE \in \HTEspace_q} \HTEnorm{\HTE - \HTEtrue}$.

\begin{theorem}[Martingale R-learner]\label{thm:MRL}
Under Assumptions~\ref{assume:causal}~and~\ref{assume:MRL}, for Martingale R-learner asymptotically solving the estimating equation~\eqref{eq:EE_basis} with $\hatscore = o(1)$,
the estimation error of HTE
has the rate
\begin{equation*}
   \MSEnorm{\hat{\HTE}-\HTEtrue}  =   \bgOp{\MSEnorm{\HTEq - \HTEtrue}
    + \frac{\basU}{\sqrt{n}} +\hatscore+ \rpsnorm{\rpsh-\rpstrue}^2+ \Lamnorm{\mLamh-\mLamtrue}^2 }.
\end{equation*}
\end{theorem}

The detailed proof is provided in Section~\ref{app:proof_MRL} of the Supplementary Materials.
%We first verify that the Assumption~\ref{assume:general}
%are satisfied under Assumptions~\ref{assume:causal} and~\ref{assume:rate_MRL}, which has been partially addressed in Lemma \ref{lemma:neyman_mrl}.
%Then, we apply the general Theorem \ref{thm:general}.
%Finally, we calculate the rate of $\E_{\inew}\left\{\psi\left(\hat{\HTE},\NMk;\datcmb_{\inew} \right)[\hat{\HTE}-\HTEtrue] \right\}$ under Assumption~\ref{assume:MRL} to specify the rate in the conclusion.
The first two terms $\MSEnorm{\HTEq - \HTEtrue}
    + \basU/ \sqrt{n}$
represent the bias-variance trade off in the estimation of $\HTE$. This part of estimation error is always present in spline regressions \citep{donald1994series}.
The last two terms are the squared estimation errors for the nuisance parameters $\rps$ and $\Lambda_{m}$.
We specify in the following corollary when the Martingale R-learner achieves the oracle property with spline basis \eqref{def:basis} and spline regressions for $\rps$ and $\Lambda_{m}$. Recall $p$ is the dimension of $\bZ$.

\begin{assumption}[Polynomial spline models]\label{assume:spline}
%\lily{we said natural or B-splines, are these special cases of polynomial spline?} \hj{Theoretical papers tend to use the general ``polynomial spline'' and give natural/B-splines as examples.}
Let $\rate[\HTE],\rate[\lambda],\rate[\rps] \in (0,+\infty)$ be absolute constants.
\begin{enumerate}[label = (\roman*), ref = \ref{assume:spline}-(\roman*)]
   \item\label{assume:spline-rates} \textbf{Standard estimation rates}.
   Under smoothness parameters $\rate[\rps]$ and $\rate[\lambda]$ for nuisance parameters, the estimated polynomial regressions with no jumps ($\mlamhn = 0$)
   achieve the standard rate:
   %\hj{Defined in Assumption~\ref{assume:cumhaz-TV}}
   $$
   \HTEnorm{\rpsh-\rpstrue} = n^{-\frac{\rate[\rps]}{2\cdot \rate[\rps] + p}}, \;
   \HTEnorm{\mlamht-\mlamtrue} = n^{-\frac{\rate[\lambda]}{2\cdot \rate[\lambda] + p}}.
   $$

   \item\label{assume:spline-approx}  \textbf{Standard approximation rate}. The approximation of HTE in the subspace $\HTEspace_q$ with $\HTEnorm{\basis_j} = 1$, $\Unifnorm{\basis_j} = O\left(q^{1/2}\right)$  and smoothness parameter $\rate[\HTE]$, achieves the standard rate
   $\HTEnorm{\HTEq - \HTEtrue} = O\left(q^{-\rate[\HTE]/p}\right)$.

    % \item \label{assume:unifZ} \textbf{Bounded covariates}.
    % The covariates $\bZ_i$ is uniformly bounded,
    % $\P\left(\Unifnorm{\bZ_i} \le \const{8}\right) = 1$.
    % \item \label{assume:holder} \textbf{H\"{o}lder smoothness}. The true HTE model~\eqref{def:HTE},
    % marginal hazard model~\eqref{def:marg_mart}
    % and the models involved in risk-set propensity score~\eqref{def:risk-ps} are all H\"{o}lder continuous (see Definition~\ref{adef:holder} in Section~\ref{app:defs} of Supplementary Materials) with smoothness parameters
    % $\rate[\HTE]$, $\rate[\lambda]$ and $\rate[\rps]$, respectively.
    \item \label{assume:spline-oracle} \textbf{Smoothness for oracle property}. $\min\{\rate[\rps],\rate[\lambda]\} \ge \rate[\HTE] p /(2\rate[\HTE]+2p)$.
\end{enumerate}
\end{assumption}

The established approximation rate and estimation rates for polynomial spline regressions in Assumption~\ref{assume:spline-rates}
are typically achieved under H\"{o}lder smooth models over bounded covariates \citep{kooperberg1995l2,newey1997convergence}. The uniformly bounded subsets of these H\"{o}lder smooth function classes are well-known Donsker classes required by Assumption~\ref{assume:donsker} \cite[Example 2.10.25]{vandervaart1996weak}.

The optimal rates for non-parametric estimation of H\"{o}lder smooth models has been established for standard and hazard regressions \citep{stone1982optimal,kooperberg1995l2}. In both work, the optimal $L_2$ rate with covariate dimension $p$ and smoothness parameter $\rate[\HTE]$ is $n^{-\rate[\HTE]/(p+2\rate[\HTE])}$. In Corollary~\ref{cor:MRL_Bspline} we show that through the Martingale R-learner, we achieved the same rate.
    % As the estimation of the structural HTE parameter under model~\eqref{def:SCFTM} is naturally harder than standard regression due to lack of directly observable contrasts in potential outcomes, we expect\lily{?} the optimal $L_2$ rate should be no less than $n^{-\rate[\HTE]/(p+2\rate[\HTE])}$, which suggests the optimality of Martingale R-learner.

\begin{corollary}[Polynomial spline]\label{cor:MRL_Bspline}
    Under Assumptions~\ref{assume:causal},~\ref{assume:MRL}~and~\ref{assume:spline}, with (a) polynomial spline $\Vbasis$ of order greater than $\rate[\HTE]$ and dimension $q = n^{1/(1+2\rate[\HTE]/p)}$, and (b) no regularization or a ridge penalty $\pennorm{\cdot}=\MSEnorm{\cdot}^2$ with $\eta = \bgOp{n^{-\rate[\HTE]/(p+2\rate[\HTE])}}$, the estimation error of Martingale R-learner is
    $
    \MSEnorm{\hat{\HTE}-\HTEtrue} = \bgOp{n^{-\rate[\HTE]/(p+2\rate[\HTE])}}
    $.
\end{corollary}
The proof of Corollary~\ref{cor:MRL_Bspline} is in the Section~\ref{app:proof_Bspline} of the Supplementary Materials.
%is mostly applying the  in Assumption~\ref{assume:spline} to Theorem \ref{thm:MRL}.
The resulting estimation rate for $\tau$ in Corollary~\ref{cor:MRL_Bspline} only involves the smoothness of HTE, $\rate[\HTE]$, indicating the oracle property.
Because $\min\{\rate[\rps],\rate[\lambda]\} \ge \rate[\HTE]$ implies the Assumption~\ref{assume:spline-oracle},
the oracle property can be achieve as long as
the nuisance models are H\"{o}lder smooth with
at least the half smoothness of HTE by inequality $\rate[\HTE]/2 \ge \rate[\HTE] p /(2\rate[\HTE]+2p)$.
This mild smoothness requirement reveals one essential
advantage of the Martingale R-learner compared to deriving HTE directly from S-learners or T-learners for hazard regressions, which would be with rate $n^{-\rate[\lambda]/(p+2\rate[\lambda])}$.
When the outcome survival model is much more complex than the HTE \citep{kennedy2023towards}, reflected in less smoothness of outcome survival model compared to HTE, $\rate[\lambda] \ll \rate[\HTE]$, the improvement from the $n^{-\rate[\lambda]/(p+2\rate[\lambda])}$ rate of the S-learners or the T-learners to the $n^{-\rate[\HTE]/(p+2\rate[\HTE])}$ rate of Martingale R-learner would be substantial.

\csection{Simulation Experiments}\label{sec:simulation}

We assessed the finite sample performance of the Martingale R-learner in simulation studies.
Based on the simulation studies from related literatures \citep{nie2021quasioracle,foster2023orthogonal},
we designed the following scenarios, specified
by the dimension of covariate $p$, baseline propensity score $\ps(\bZ_i)$, HTE $\HTE(t,\bZ_i)$ and the hazard function of control group $\lambda(t;\bZ_i,0)$.
%\hj{$p$ was never specified.}
\begin{enumerate}[label = \textbf{\Alph*}., ref = \Alph*]
    \item\label{sim:A} $p=1$, $\ps(Z_i) = 0.5$, $\HTE(t,Z_i) = Z_i+Z_i^2 +t +t^2$, $\lambda(t;Z_i,0) = 0.5$.
    \item\label{sim:B} $p= 3$, $\ps(\bZ_i) = \expit(Z_{i,1}+Z_{i,2}+Z_{i,3})$, $\HTE(t,\bZ_i) = \sin(Z_{i,1}+Z_{i,2}+Z_{i,3})+\sin(t)+4$, $\lambda(t;\bZ_i,0) = 4 + Z_{i,1}$.
    \item\label{sim:C} $p= 3$, $\ps(\bZ_i) = \expit(Z_{i,1})$, $\HTE(t,\bZ_i) = 4+\log\{1+\exp(Z_{i,2}+Z_{i,3})\}+t$, $\lambda(t;\bZ_i,0) = 3 + Z_{i,1}$.
    \item\label{sim:D} $p= 3$, $\ps(\bZ_i) = \expit(-Z_{i,1}-Z_{i,2}-Z_{i,3})$, $\HTE(t,\bZ_i) = 6$, $\lambda(t;\bZ_i,0) = \log\{1+\exp(Z_{i,1}+Z_{i,2})\}+t^2$.
    \item\label{sim:E} $p= 5$, $\ps(\bZ_i) = \expit(-Z_{i,1}-Z_{i,2}-Z_{i,3})$, $\HTE(t,\bZ_i) = \log\{6+Z_{i,3}+Z_{i,4}\}+\sin(Z_{i,5}) +\sin(t)$, $\lambda(t;\bZ_i,0) = 3 + Z_{i,1}$.

\end{enumerate}
Here $\expit(x) = 1/(1+e^{-x})$ is the link function
for logistic regression.
Across all scenarios, the covariates $\bZ_i$ are generated from independent uniform distribution $\Unif(-2,2)$.
In these scenarios, we defined polynomial HTEs and randomized treatment in Scenario \ref{sim:A},  complex HTEs with trigonometric and logarithm-exponential functions and logistic regression propensity scores in Scenarios \ref{sim:B} -- \ref{sim:E}.
We considered time-invariant control group hazard functions in Scenarios \ref{sim:A} -- \ref{sim:C} and \ref{sim:E}, and set a time-varying hazard function in Scenarios \ref{sim:D}.
The associations between baseline propensity scores and control group hazard functions were positive in Scenarios \ref{sim:B} and \ref{sim:C}
and negative in Scenarios \ref{sim:D} and \ref{sim:E}.
The censoring time $C_i$ are generated from $\Unif(0,\cmax)$ with $\cmax$ chosen
to produce $15\%$ censoring rate.

%\hj{No details about the implementation?}
In all the simulation studies, we implemented and deployed the Martingale R-learner (\textbf{MRL}) as specified in the following.
% We estimated the marginal survival model by regressing the observed survival outcome on $\bZ_i$, using hazard regression with linear splines \citep{kooperberg1995hazard, stone1997polynomial} \Y{$S_m(t \mid Z)
% =
% \{1-\ps(Z)\}S_0(t \mid Z)
% +
% \ps(Z)S_1(t \mid Z)$}.
% %\Y{The two references in the hare function description}.
We estimated the (baseline) propensity scores $\psk(\bZ_i)$ by regressing $\Trt_i$ on $\bZ_i$ using a generalized additive logistic regression model \citep{hastie1986generalized} and derived the conditional survival function $\Survk(t;\bZ_i, \Trt_i)$ from hazard regression with  the survival outcome on linear splines of $\Trt_i$ and $\bZ_i$   \citep{kooperberg1995hazard, stone1997polynomial}. We obtained $\rpsk$ and $\mLamk$ by assembling $\psk$ and $\Survk$ through \eqref{eq:risk-ps-alt} and \eqref{eq:margHaz-alt}, respectively.
For all simulation scenarios, we approximated the HTE using natural spline basis expansions for time and continuous baseline covariates, together with their pairwise interactions. We used the same degrees of freedom for time and for each continuous covariate, with candidate values $\{5,6,7,8,9\}$. The spline dimension and ridge penalty parameter were selected by cross-validation using the pseudo least-squares criterion.
As the benchmark, we also
implemented the S-learner using hazard regression with linear splines (\textbf{SL-HARE}) and the T-learner using the same method with two separate models for the control and treated groups (\textbf{TL-HARE}).
To assess the oracle property, we also implemented the Martingale R-learner with oracle nuisance models.
%oracle risk-set propensity scores (\textbf{Oracle-$\rho$}),
%oracle marginal survival models (\textbf{Oracle-$\Lambda_m$}) or both (\textbf{Oracle-both}).

For each scenario, we considered three difference sample sizes $n=500, 750, 1000$ to evaluate the trajectory of estimation performance with growing sample sizes.
We repeated the experiments 500 times and report the mean squared error (MSE) evaluated over an independent testing data $\{(X_{*,i},\bZ_{*,i}):i=1,\dots,n\}$, which is same across repeats to eliminate uncertainty,
$$
\MSE{\hat{\HTE},\HTEtrue} = N^{-1}\sum_{i=1}^N \left\{\hat{\HTE}(X_{*,i},\bZ_{*,i})-\HTEtrue(X_{*,i},\bZ_{*,i})\right\}^2.
$$

\begin{figure}[htbp]
    \centering
    \includegraphics[width=16cm]{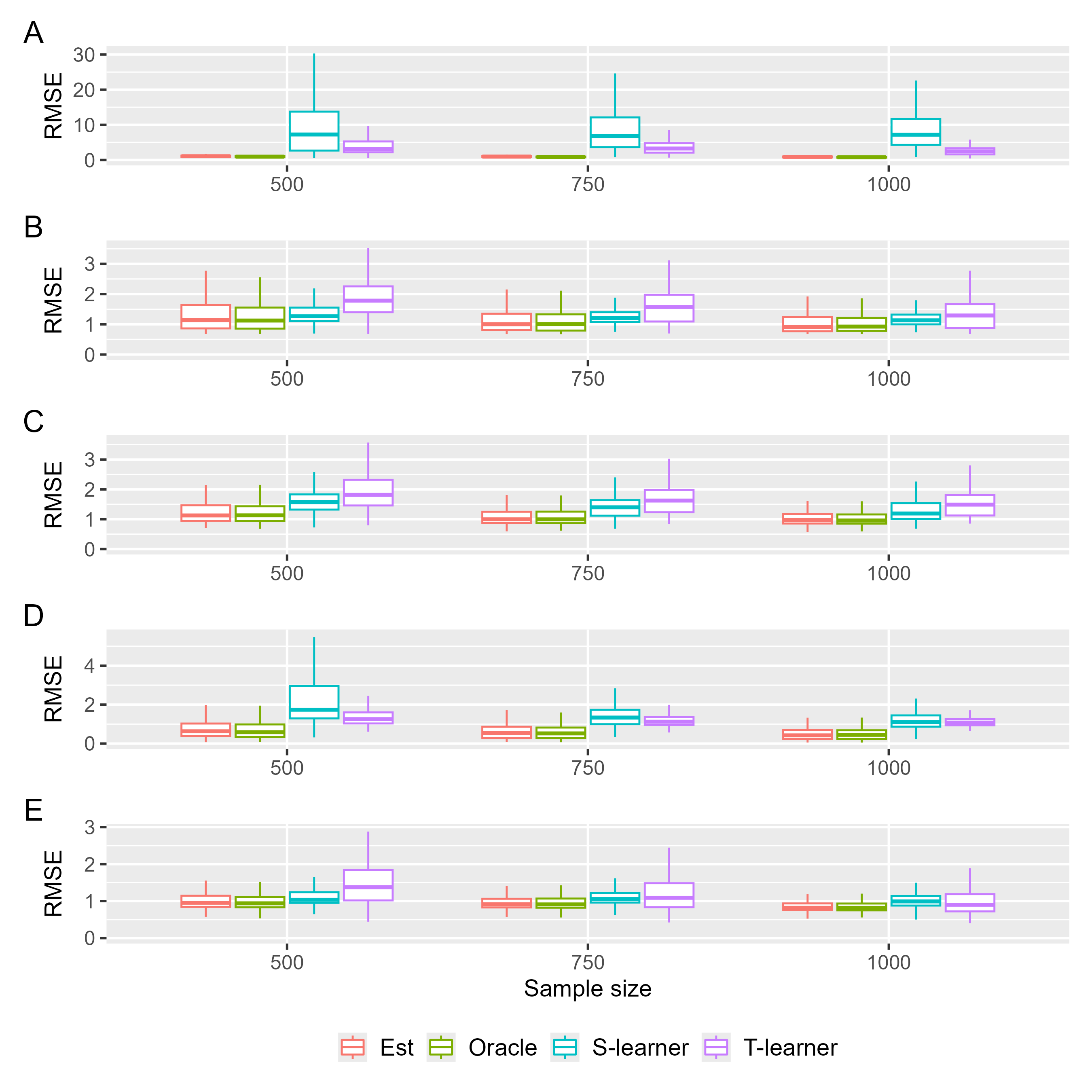}
    \caption{Distribution of root MSE under scenarios A - E.
    %\hj{Might want to fit Figures 1 and 2 into one page. Reconsider the legends based on my new writing. Consider adding the benchmarks. Scenario E looks wrong with oracle worse than others.}
    }
    \label{fig:sim}
\end{figure}

We reported the simulation results in Figure \ref{fig:sim}. %\hj{Comment on the comparison of MRL vs SL or TL.}
Across all five simulation scenarios and all sample sizes considered, the Martingale R-learner with estimated nuisance functions achieved lower median RMSE than the S- and T-learners. In Scenarios \ref{sim:A} and \ref{sim:D} where the HTE was set as  polynomial functions favorable for spline methods, Martingale R-learner significantly outperformed the benchmark S- and T-learners exhibited with much smaller RMSE and variability.
In Scenarios \ref{sim:B}, \ref{sim:C}, and \ref{sim:E} where the HTE was complex, the Martingale R-learner still demonstrated were relatively moderate advantage over benchmark S- and T-learners. As expected, the oracle Martingale R-learner generally achieved the lowest RMSE with the extra information of true nuisance models. The gap between the estimated nuisance and oracle nuisance learners tended to decrease as the sample size increased (see also Figure~\ref{Sfig: MSE ratio} in the Supplementary Materials), indicating the oracle property of the Martingale R-learner that the effect of nuisance estimation error diminished with the increasing sample size. RMSE also generally decreased with the sample size for all methods, while the relative advantage of the Martingale R-learner remained evident.

%\Y{RMSE, show the method names, change the legend labels, show ratio plots}

\csection{Application}\label{sec:real_data}

We applied the Martingale R-learner implemented as in Section~\ref{sec:simulation} to the motivating study on mid-life heavy drinking on the cognitive-impairment-free-survival.
%to data from the Honolulu Heart Program (HHP) and its successor, the Honolulu Asia Aging Study (HAAS) with 3,734 participants \citep{peila2001joint}. The HHP enrolled a cohort of men with Japanese ancestry born between 1900 and 1919 who were living on the island of Oahu, Hawaii, and followed them for cardiovascular outcomes. In the HAAS established in 1991, the focus shifted to late-life cognitive outcomes, including incident dementia and cognitive impairment.
Cognitive function was assessed repeatedly at HAAS examinations using the Cognitive Abilities Screening Instrument (CASI), a 0--100 scale covering attention, concentration, orientation, memory, language, visual construction, fluency, abstraction, and judgment. Death was ascertained through study surveillance and follow-up contacts, with death dates obtained from death certificates.
From the baseline at study Exam 4, we define the cognitive-impairment-free-survival outcome as as the time from baseline to the first HAAS examination at which the participant had a CASI score below 74, or mortality, whichever comes first.
Participants alive with all CASI scores above 74 were censored at their last HAAS examination with available CASI assessment.
The mid-life heavy drinking status was defined as $\geq$ 14 drinks/week at either study Exam 1 or Exam 3. We considered covariates including the baseline age at Exam 4, education ($\le$ 12 years or $>$ 12 years) assessed at the start of study Exam 1, \apoe[]  genotype (\apoe[4]+ or \apoe[4]-) and grip strength (kilograms of force, defined as the maximum over six grip measurements at study Exam 4, consisting of three dominant hand and three opposite hand measurements. After excluding observations with missing data in any of exposure, outcomes or covariates, the analytic data set contained 1874 participants. Baseline characteristics are summarized in Table \ref{Stab:haas_summary} (Section~\ref{app:haasKM} of Supplementary Materials).

\begin{figure}[htbp]
\centering
\includegraphics[width=0.49\textwidth]{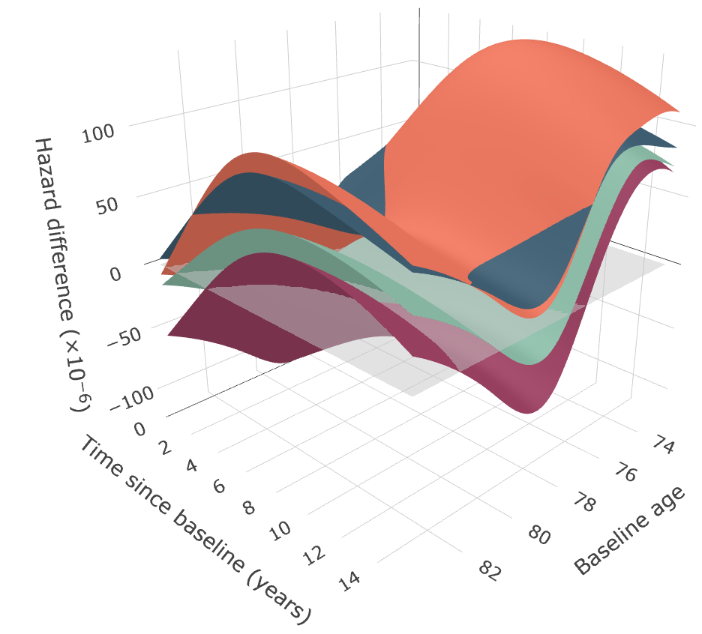}
\includegraphics[width=0.49\textwidth]{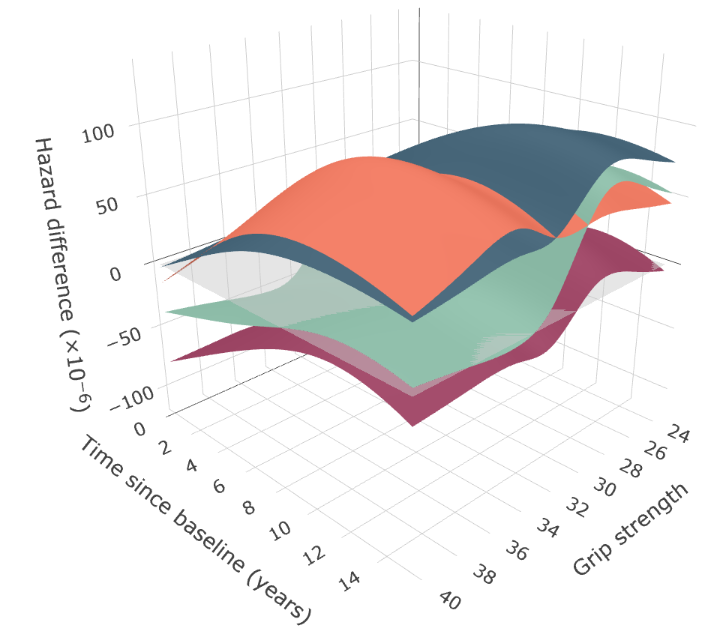}

\vspace{0.5em}

\includegraphics[width=0.49\textwidth]{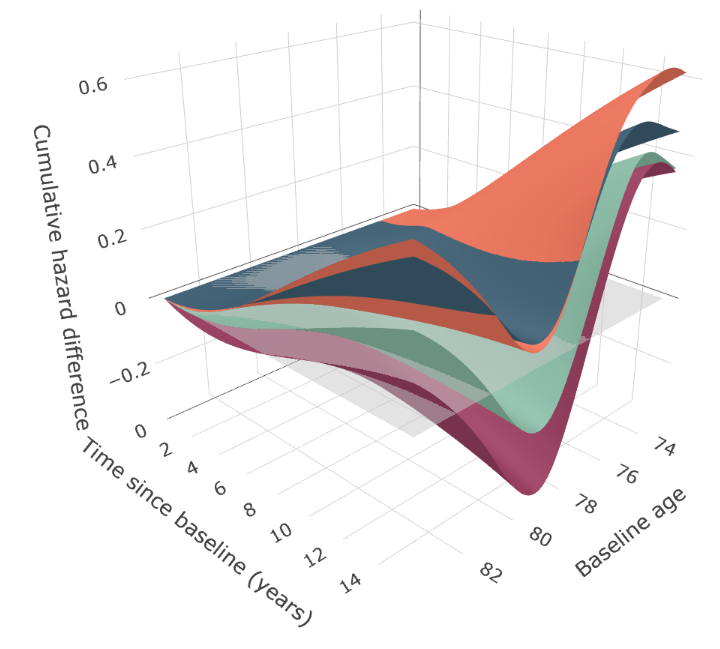}
\includegraphics[width=0.49\textwidth]{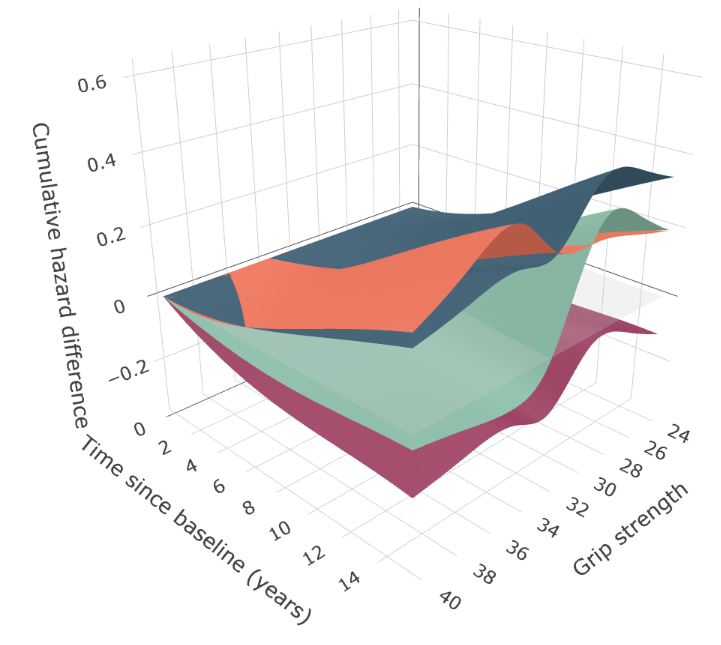}

\vspace{0.5em}

\includegraphics[width=0.5\textwidth]{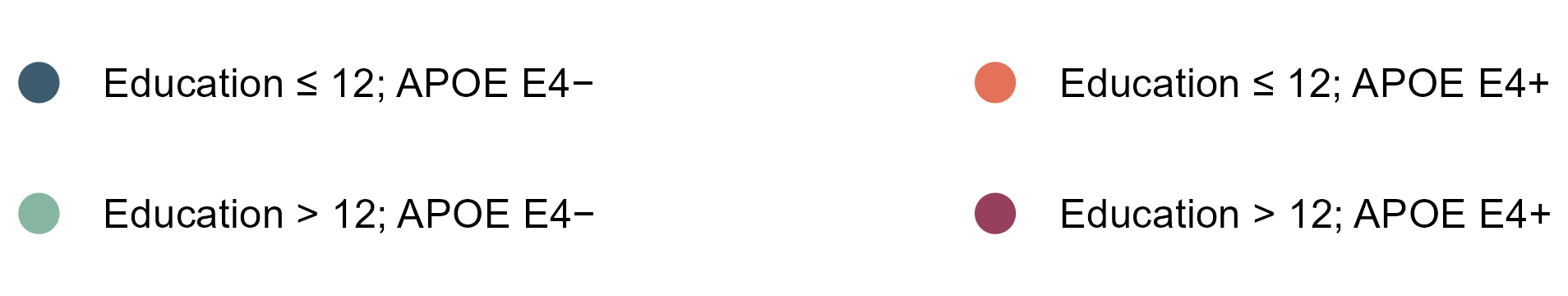}

\caption{
Estimated heterogeneous treatment effect surfaces for midlife heavy drinking in the HAAS application. The top row shows the infinitesimal treatment effect, and the bottom row shows the cumulative hazard difference. The left column varies baseline age and time, with grip strength fixed at its mean; the right column varies grip strength and time, with baseline age fixed at its mean. Surfaces are shown separately by education and APOE.
}
\label{fig:haas}
\end{figure}

% The HAAS study has a big cohort with almost 2,000 subjects, thus heterogeneity across different individuals cannot be ignored. In addition, the study had a very long follow-up time, then in order to account for that the effect may change over time, considering time as a variable will be an appropriate approach. These two issues require a method for estimating time-varying heterogeneous treatment effects for time-to-event outcomes, and our martingale R-learner is applied to deal with these two concerns.
Figure \ref{fig:haas} shows the estimated HTE surfaces varying by time and baseline age or grip strength for four strata defined by education and \apoe[] genotype, where age or grip strength is fixed at its average value. The estimated HTE surfaces display non-linear features across time and baseline age or grip strength. The HTE is positive for most values of covariates, suggesting that mid-life heavy drinking is associated with higher risks of cognitive impairment.
Moreover, we observed that the harm of mid-life heavy drinking is smaller for groups with better education.
As better education were found to associate with better cognitive outcomes \citep{meng2012education},
our finding seems to be consistent with previous analysis.
We presented additional analysis results in Section~\ref{app:haasKM} of Supplementary Materials including visualization of the estimated HTE surfaces from other angles and subgroup descriptive analyses to support the findings and estimated HTE.

%\lilyemph{please plot $\tau$ also not cumulative; like I said should plot both}
%\lily{add zero plane}
%\lily{can you add lines on the surface to see effects over t for fixed z? Maybe you can turn each surface into like a net?} \hj{Change legent to \apoe[4]+ or \apoe[4]-.\lily{it would be APOE E4+ etc} Also information about the reference value of grip on the left and age on the right (mean).}

%\hj{Add the stratified Kaplan-Meier for treated vs control in HTE subgroups.}

\csection{Discussion}\label{sec:discuss}

In this work, we developed the Martingale R-learner for estimating
time-varying HTE on right censored time-to-event outcomes.
With the oracle property, the Martingale R-learner can produce faster estimation rate for HTE with flexible machine learning methods for nuisance models with slower rates and can attain standard optimal nonparametric estimation rate for smooth HTE.
To formalize our methodology and theory, we introduced the functional score framework that extended traditional estimating equations to nonparametric parameters and generalized the concept of Neyman orthogonality accordingly. Unlike the traditional problem with finite dimensional parameter space, the connection between loss functions and functional scores as explained in Remark~\ref{remark:neyman_fs} does not hold universally with infinite dimensional parameters.
The differentiation was shown to be significantly more restrictive \citep{luedtke2024one}, and the existence of anti-derivative loss with a given derivative functional is a topic in quantum mechanics \citep{feynman2010quantum}.

While
various regression or classification based methods have previously been applied for learning HTE
without explicit involvement of the counterfactual alternative \citep{willke2012concepts}, under the potential outcome framework causal inference models characterize the HTE through structural nested models and marginal structural models \citep{vansteelandt2014structural,robins2000marginal_epi}.
In this work, we adopt the structural nested model framework for time-to-event outcomes,
namely, the structural nested cumulative failure time models  \citep{robins2008estimation, picciotto2012structural}, and define the HTE through the time-varying survival ratio of potential time-to-event outcomes between the two treatment options in comparison conditionally on all confounders.
While the distinction between these two classes of models originally stemmed from the different strategy for handling time-varying confounders in the analysis of time-varying treatment \citep{robins2000marginal},
the idea of ``marginalizing out'' some confounding variables from effect modification motivated the marginal structural models for point treatment \cite[Section 12.5]{HernanRobinsWhatif}. In our current study, we focused on the problem of the HTE with a point treatment applied before the start of follow-up without any \emph{a priori} selection for effect modifiers from candidate confounders. Extension to R-learner marginal structural models can be derived from generalizing a proper estimating equation through the functional score framework.

Beyond the implemented Martingale R-learner with hazard regressions, other machine learning methods for survival outcomes can also be considered, such as high-dimensional regressions, random forests or boosting \citep{gaiffas2012high,ishwaran2008random,binder2008allowing}.
While we focused on the estimation of the infinite dimensional HTE,
our theoretical analysis explicitly identified the variance and bias components within the estimation error, which may prepare the statistical inference for its low dimensional projections with under-smoothing \cite[Chapter 6]{ruppert2003semiparametric}.
The HAAS study is potentially subject to left truncation as patients died before study Exam 4 were excluded in the analysis.
Future study should revisit the analysis after incorporating the general strategy to correct for left-truncation as laid out in \citet{wang2024learning}.

\csection{Acknowledgment}

Jue Hou and Yuchen Qi have equal contribution as co-first authors.
The study codes and supplementary materials of technical details and theoretical proofs are not made available for the preprint version.

  \bibliography{Time-varying-HTE-fix}

\end{document}